%% file: main.tex
\documentclass{article}
\usepackage{spconf,graphicx}

\usepackage{cite}
\usepackage{amsmath,amssymb,amsfonts}
\usepackage{algorithmic}
\usepackage{adjustbox} 
\usepackage{textcomp}
\usepackage{xcolor}
\def\BibTeX{{\rm B\kern-.05em{\sc i\kern-.025em b}\kern-.08em
    T\kern-.1667em\lower.7ex\hbox{E}\kern-.125emX}}
\usepackage{subcaption}

\usepackage[implicit]{hyperref}

\usepackage{url}
\usepackage{lipsum}
\usepackage{booktabs}

\usepackage{microtype}

\input{notations}

\title{Neural Steerer: 
Novel Steering Vector Synthesis \\
with a Causal Neural Field
over Frequency and Direction}

\name{
    \kern-0.6em
    Diego Di Carlo$^{1,2}$ \kern0.4em
    Aditya Arie Nugraha$^{1,2}$ \kern0.4em
    Mathieu Fontaine$^{3,1}$ \kern0.4em
    Yoshiaki Bando$^{4,1}$ \kern0.4em
    Kazuyoshi Yoshii$^{2,1}$
    \thanks{
        This work was supported by 
        ANR Project SAROUMANE (ANR-22-CE23-0011) and Hi! Paris Project MASTER-AI, JST PRESTO no.~JPMJPR20CB,
        and JSPS KAKENHI nos.~JP20H00602, JP21H03572, JP23K16912, JP23K16913.
        \\Code available at \url{https://github.com/Chutlhu/nsteerer}.
    }
}
\address{
    $^{1}$\,Center for Advanced Intelligence Project (AIP), RIKEN, Japan \\
    $^{2}$\,Graduate School of Informatics, Kyoto University, Japan \\
    $^{3}$\,LTCI, Télécom Paris, Institut Polytechnique de Paris, France \\
    $^{4}$\,National Institute of Advanced Industrial Science and Technology (AIST), Japan \\
}

\begin{document}
\ninept
\maketitle
\begin{abstract}
We address the problem of accurately interpolating measured anechoic steering vectors with a deep learning framework called the \textit{neural field}. 
This task plays a pivotal role in reducing the resource-intensive measurements required for precise sound source separation and localization, essential as the front-end of speech recognition. 
Classical approaches to interpolation rely on linear weighting of nearby measurements in space on a fixed, discrete set of frequencies.
Drawing inspiration from the success of neural fields for novel view synthesis in computer vision, we introduce the \textit{neural steerer}, a continuous complex-valued function that takes both frequency and direction as input and produces the corresponding steering vector.
Importantly, it incorporates inter-channel phase difference information and a regularization term enforcing filter causality, essential for accurate steering vector modeling.
Our experiments, conducted using a dataset of real measured steering vectors, demonstrate the effectiveness of our resolution-free model in interpolating such measurements.

\end{abstract}
\section{Introduction}
Steering vectors are a fundamental concept in multichannel audio signal processing as they describe the acoustic relationship between a sound source and a set of microphones in anechoic settings.
It serves as a core component of speech enhancement~\cite{donley2021easycom}, source separation~\cite{sekiguchi2020fastmnmf} and localization~\cite{schmidt1986multiple}, and acoustic channel estimation~\cite{annibale2012geometric}.
Hence, their accurate representation is fundamental for robust sound analysis and to achieve realistic rendering.
In most actual applications, steering vectors are computed based on algebraic anechoic models or estimated by measuring head-related transfer functions (HRTFs), directivity profiles of a microphone array, and acoustic transfer functions. 

Algebraic steering vectors analytically encode the direct propagation over space and frequency, typically as a function of sound incident angle. In real scenarios, however, such an algebraic model is limited by several impairments and often replaced with measured general filters~\cite{gannot2001signal}. Extended formulations include models for directivity and filtering as well as sound interaction with the receiver, such as occlusion, diffraction, and scattering. In the hearing aid applications, the steering vectors encode HRTF that capture the effects of the user's pinnae, head, and torso. In order to model all the effects at each space and frequency, one may use dedicated simulators, which suffer from significant realism-computational complexity trade-offs~\cite{argentieri2015survey}.
Alternatively, steering vectors can be measured in dedicated facilities~\cite{donley2021easycom}.
However, measuring them at high spatial resolution is cumbersome due to the cost and setup complexity, if not unfeasible. 

\begin{figure}[!t]
    \centering
    \includegraphics[trim={0 23 210 5},clip,width=\linewidth]{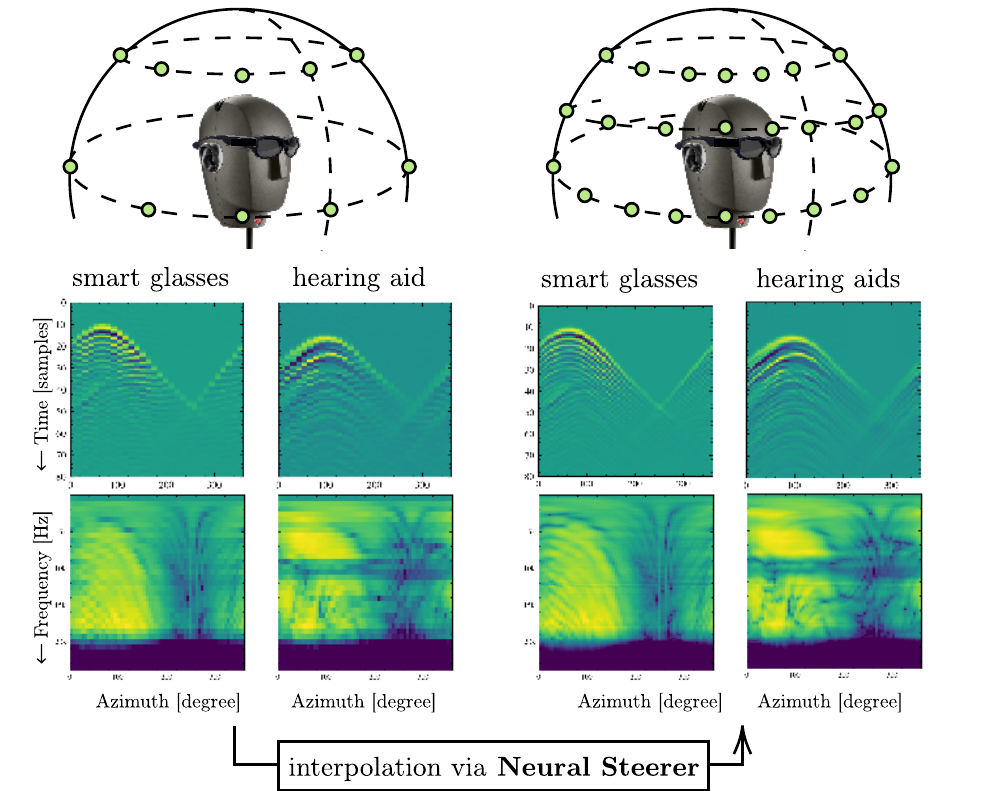}
    \caption{Schematic representation of the proposed Neural Steerer for measured anechoic steering vectors interpolation.}
    \label{fig:sv_interpolation}
    \vspace{-7mm}
\end{figure} 

Reliable data-driven approaches for steering vector and HRTF interpolation have gained much attention over the last decade~\cite{pulkki1997virtual,freeland2004interpolation,zotkin2009regularized}.
As shown in Figure \ref{fig:sv_interpolation}, These methods use measurements on a coarse spatial grid as a basis and then interpolate them to obtain measurement estimates at new locations, typically on a finer grid.
However, such methods suffer from two main drawbacks.
First, they assume that the target quantity undergoes a polynomial transformation of some basis function at the desired spatial location, which may not accurately represent reality and introduce artifacts.
Second, all these methods operate interpolation only in the spatial domain, leaving the frequency (or temporal) grid resolution fixed, under-modeling the target quantity.

Recent works on deep learning showed that \textit{coordinate-based} architectures are able to encode signals continuously over space and time~\cite{tancik2020fourier,sitzmann2020implicit,xie2022neural}.
These methods, also called \textit{neural fields} (NFs), can be trained easily using data-fit and model-based loss functions in order to interpolate and super-resolve target quantity at an arbitrary resolution.
Their outstanding applications include novel view synthesis from sparse 2D images~\cite{mildenhall2020nerf}, image super-resolution~\cite{sitzmann2020implicit}, and animation of human bodies~\cite{gafni2021dynamic}.
Unlike on-grid discrete measurements at a given resolution, the memory complexity of an NF scales with the data complexity and not with the desired resolution.
In addition, being over-parameterized modular network models, NFs are effective for regressing ill-posed problems under appropriate regularization.
However, in practice, they tend to overfit, and their resolution capabilities are limited by network capacity and training schemes~\cite{xie2022neural}.

Interestingly, \textit{physics-informed neural networks} (PINNs)~\cite{raissi2019physics,karniadakis2021physics,wang2021understanding} are similar models proposed in the field of computational physics. Here, partial differential equations (PDEs) are used as regularization terms evaluating coordinates at arbitrary resolution exploiting automatic differentiation. The resulting NFs are likely to be a physically consistent continuous function defined over the whole input domain and less prone to overfit to noisy or sparse input observations. However, these networks are required to be differentiable with respect to the input variables used in the PDEs, which is not the case for available audio-based NF models returning a discrete set of frequencies.

In this work, we address the limitation of available discrete steering vector interpolation methods by leveraging recent advances in neural fields. The proposed method models both the magnitude and the phase of steering vectors (related to HRTFs) as a complex-valued field defined continuously over space and frequency.
Based on the theoretical foundation of signal processing, we propose a novel regularizer that enforces the estimated filter to be causal, leading to more accurate filters with respect to standard approaches. Finally, thanks to its continuous formulation in space and time, the proposed approach can be used in the framework of PINNs evaluating the wave-equation terms with respect to the model inputs.

\section{Related Work}

Given a set of steering vector measurements in the frequency or time domain, a simple way to increase the spatial resolution is to obtain an estimate for a new position by weighted averaging known measurements from multiple surrounding positions.
Methods of this form are extensions of polynomial interpolation where data are projected onto ad-hoc basis functions (e.g., spherical harmonics or spatial characteristic function) whose coefficients are then interpolated at desired positions~\cite{pulkki1997virtual,freeland2004interpolation,zotkin2009regularized}.
The well-known limitations of these approaches are the requirement of near-uniform sampling for the training data and the difficulty of conditioning the optimization on ad-hoc prior information. Also, the performances are related to the choice of the basis function and their interpolating algorithms.

In audio processing, NFs have recently been proposed for acoustic impulse response (AIR) estimation and interpolation ~\cite{richard2022deep}, for HRTF magnitude encoding over arbitrary spherical coordinates~\cite{zhang2022hrtf,lee2022global}, and binaural rendering~\cite{lee2022neural,luo2022learning,su2022inras}.
Interestingly, in \cite{lee2022global}, estimation of AIRs is performed incorporating explicitly the algebraic model of early sound propagation. However, these approaches treat frequency (time) as a discrete quantity corresponding to the output dimension of neural networks, which limits the model's applicability.

By explicitating frequency (or time) as a network's input variable, the model can be trained within the PINN framework, leveraging the wave equation to enforce physics coherence. This approach has been used to recover AIRs at unseen locations in the time domain~\cite{pezzoli2023implicit} and to spatially super-resolve complex HRTF for a single given frequency bin~\cite{ma2023physics}. Besides the promising results, no guarantees are given on the shape of estimated filters, which may feature, for instance, anti-causal components.
Moreover, it is worth noting that, apart from~\cite{lee2022neural,ma2023physics}, the phase components of the steering vector are ignored. This limitation can significantly affect spatial processing and rendering, where phase information is crucial for accurate processing.

\section{Proposed Method}
It is reasonable to start with the theoretical computation of steering vectors. Let us consider an anechoic steering vector that encodes the direct signal propagation from the $j$-th source located at position $\srcPos_j$ to an $I$-microphone array centered at $\refPos$ as a function of the incident direction of arrival (DoA) represented by azimuth $\az_j$ and elevation $\el_j$.

Following the far-field \textit{algebraic} model in the frequency domain~\cite{vincent2018audio},
the anechoic steering vector $d_{ij}(f)$ for the $i$-th microphone located at $\micPos_i$ at frequency $f$ is expressed as
\begin{equation}\label{eq:theory_svect}
    d_{ij}(f) = \exp\left(- \frac{\jmath 2 \pi f \mathbf{n}_j^\top (\micPos_i - \refPos)}{c}\right) ,
\end{equation}
where $\mathbf{n}_j = [ \cos \az_j \cos \el_j,\sin\az_j \cos \el_j, \sin \el_j ]^\top$ is the unit-norm vector pointing to the $j$-th source, $c$ is the speed of sound,
and $\jmath$ represents the imaginary unit.

The real-world steering vector deviates from the ideal anechoic steering vector due to various filtering effects caused by complex physical phenomena such as signal propagation in the air and microphone directivity \cite{scheibler2018pyroomacoustics}. We thus formulate a steering vector $h_{ij}(f)$  as a modified version of $d_{ij}(f)$ as follows:

\begin{equation}\label{eq:atf}
    h_{ij}(f) = \underbrace{\vphantom{\mathbf{n}_j^T} \exp(- 2\pi f \jmath \tau)}_{\triangleq \, \bar{d}(f)}
    \, g^{\text{air}}_j(f) \, g^\text{mic}_{ij}(\micPos_i, f) \,  d_{ij}(f),
\end{equation}
where $g_{j}^\text{air}(f)$ represents the frequency-dependent air attenuation and the eventual source-dependent characteristics and $g^\text{mic}_{ij}(\micPos_i, f)$ represents the microphone directivity defined with its phase center at $\refPos$, which typically depends on the type and manufacturing of the microphones\cite{srivastava2023virtually}. Finally, a global delay term $\bar{d}(f)$ accounts for the global constant time offset $\tau$, often presented in real-world measurements.

\subsection{Neural Steerer}
The family of steering vectors can be represented by the functional:
\begin{equation}
    \arraycolsep=0.1pt\def\arraystretch{1.0}
\begin{array}{lcccrc}
\model: \; & \mathbb{S}^2& \times\; & \mathbb{R}_{+} & \to & \mathbb{C}\\
 & ((\az_{j},  \el_{j}),& & f) & \mapsto & \; \refATF,
\end{array}\label{eq!model}
\end{equation}
where $\mathbb{S}^2$ is the set of polar coordinates on the unit sphere.
The dependency on $\micPos_i$ is omitted for readability as they are constant. 

We parameterize $\model$ with an NF~\cite{sitzmann2020implicit,tancik2020fourier,xie2022neural}, that, 
once trained on a finite set of observations, can evaluate any input coordinate at arbitrary resolution whose super-resolution capabilities depend on the network's inductive bias, e.g., architecture topology and regularizers. Similarly to~\cite{lee2022neural}, we model $\tau$ and $\micPos_i$ as free parameters of the model, which are optimized during training. Then, $\bar{d}$ and $d_{ij}$ can be computed algebraically from the input coordinated using~\eqref{eq:theory_svect}. Therefore, the network predict only the terms $g^{\text{air}}_j(f)$ and $g^{\text{mic}}_{ij}(\micPos_i, f)$. Such a physics-inspired formulation is helpful as it enforces the inductive bias and provides a good initialization, although the single terms may not correspond to the actual physical contributions. 

\begin{figure}[!t]
    \centering
    \includegraphics[trim={0 0 220 0},clip,width=\linewidth]{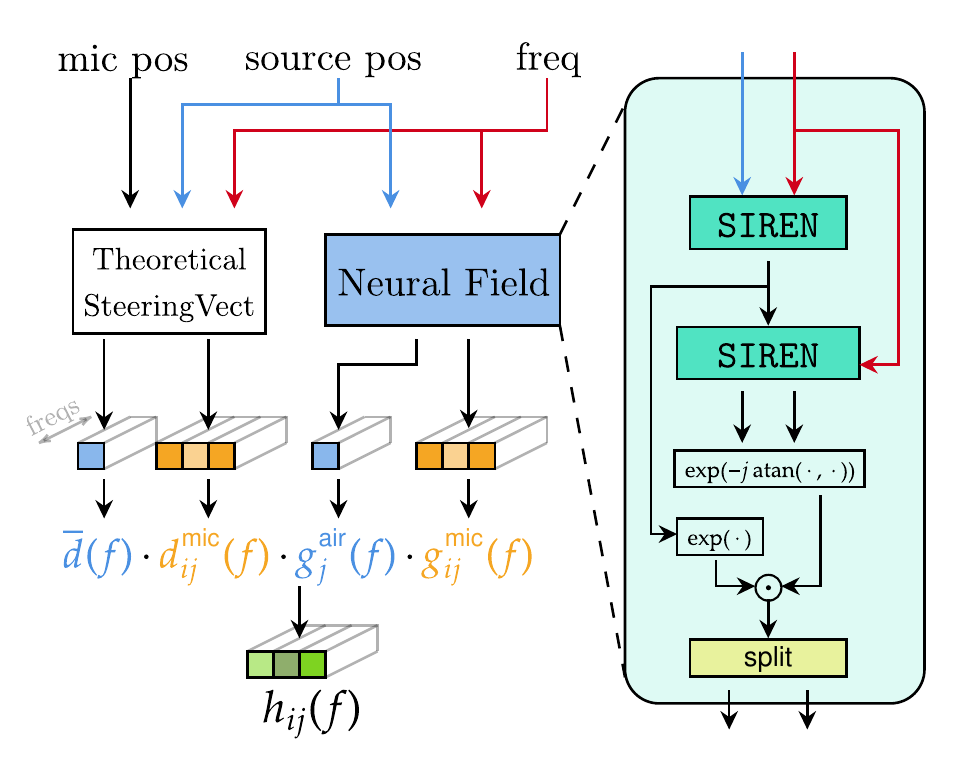}
    \caption{Illustration of the proposed Neural Steerer. Both $\odot$ and $\cdot$ denote element-wise multiplication.}
    \label{fig:model}
\end{figure}

\subsection{Network Architecture}
The proposed architecture is shown in Fig.~\ref{fig:model}.
As in~\cite{zhang2022hrtf}, a SIREN network is used, i.e., a multi-layer perceptron (MLP) with sinusoidal activations~\cite{sitzmann2020implicit}.
This model, denoted by $\mathtt{SIREN}_\text{Phase}$, takes as input the desired source DoA $(\az_j, \el_j)$ and frequency $f$ and returns the components of~\eqref{eq:atf}:
$\mathcal{G}_{j} \triangleq \left\{ 
g^{\text{air}}_j(f),
\{ g^{\text{mic}}_{ij}(\micPos_i, f) 
\}_{i=1}^I \right\}$.

Let $g^\ast$ be an entry of $\mathcal{G}_{j}$.
Instead of directly predicting the complex-valued $g^\ast$ or its real-imaginary parts, we propose to estimate its magnitude and phase implicitly.
More precisely, for each $g^\ast$, the network returns a $3$-dimensional vector $\mathbf{g}^\ast \in \mathbb{R}^3$
that is used to obtain $g^\ast$ via its magnitude and phase as
\begin{equation}\label{eq:phase}
    g^\ast= \exp \left( \{ \mathbf{g}^\ast \}_1 \right) \, \exp \left( -\jmath 2 \pi \arctan \left( \frac{ \{ \mathbf{g}^\ast\}_2 }{ \{ \mathbf{g}^\ast \}_3 } \right) \right) .
\end{equation}
Such representation handles phase wrapping, suffered by the magnitude-phase format, and enhances stability and convergence speed compared to a real-imaginary one, leading to better results.

The steering vector $h_{ij}(f)$ is then obtained as in~\eqref{eq:atf}
given $g^{\text{air}}_j(f)$ and $g^{\text{mic}}_{ij}(\micPos_i, f)$ estimated by the NF
and the theoretical steering vectors $d_{ij}^\text{mic}(f)$ and the global delay $d_j(f)$ are computed knowing the microphone positions and the offset, respectively.

Following the work in~\cite{nugraha19dgm}, we propose an alternative architecture (denoted by $\mathtt{SIREN}_{\text{Mag}\to\text{Phase}}$), where the phase estimation is conditioned on the magnitude estimation. In practice, the network comprises a cascade of two SIREN, jointly trained: the first outputs the magnitudes of the components from input coordinates, while the second reconstructs the phase based on both coordinates and magnitudes.

This proposed model considers frequencies as continuous input variables.
In this setting, hereafter referred to as \textit{continuous frequency} (CF) model, the network output $3 \times (I + 1)$ real values, where 3 is the dimension of $\mathbf{g}^\ast$ to represent complex values.
In the case of \textit{discrete frequency} (DF) modeling, the last layer of the network is modified to output $3 \times (I + 1) F$ for a given single source DoA, where $F$ is the total number of frequency regularly sampled in $[0,F_s]$, similarly to the DFT operation, where $F_s$ is the sampling frequency.

\subsection{Training Loss and Off-Grid Regularization}

The neural network is trained to return filters with low phase distortion in both frequency and time domains for a given source DoA $(\az_j, \el_j)$ and frequency $f$. Similarly to the loss proposed in~\cite{richard2021neural}, we optimize the explicit magnitude and phase errors in the frequency domain plus one explicit term time-domain $\ell_2$-loss, that is  

\begin{align}\label{eq:basicLoss}
    \mathcal{L} = 
        & \sum_{ij \in \mathcal{B}} \left( \frac{1}{| \mathcal{F} |} \sum_{f  \in \mathcal{F}} \mathcal{L}_{ijf}^\text{freq} + \frac{1}{F} \mathcal{L}_{ij}^\text{time} \right) \\
    \mathcal{L}_{ijf}^\text{freq} = 
        & \mathcal{L}_{\text{LogMag}}(\estATF(f), \refATF(f))
        + \lambda_1 \mathcal{L}_{\text{Phase}}( \estATF(f), \refATF(f)) \nonumber\\
    \mathcal{L}_{ij}^\text{time} = 
        & \lambda_2  \left\| \text{iDFT} \left( \vphantom{\hat{A}^A_A} [\estATF]_{F} \right) - \text{iDFT} \left( \vphantom{\hat{A}^A_A} [\refATF]_{F} \right) \right\|_2^2, \nonumber
\end{align}
where $\mathcal{B}$ and $\mathcal{F}$ are the sets of random training DoAs and frequencies used in a batch, respectively, with $|\mathcal{F}| < F$. $[h_{ij}]_{F}$ denotes the concatenation of $F$ elements for equally spaced frequencies in $[0, F_s]$.

In practice, as reported in~\cite{donley2022dare}, we also find the $\ell_1$ loss on the log-magnitude spectrum plus a $\ell_1$ loss on the $\cos$ and the $\sin$ of the phase component work best, denoted by $\mathcal{L}_{\text{LogMag}}$  and $\mathcal{L}_{\text{Phase}}$, respectively. 
Moreover, we found empirically that modeling \textit{continuous frequencies} (CF) is a much harder task than modeling frequencies as discrete output variables.
In particular, we found that the NF fails to capture the details across the frequency dimension, converging towards erroneous solutions.
We addressed this pathology by modifying loss functions $\mathcal{L}_{\text{LogMag}}$ and $\mathcal{L}_{\text{Phase}}$ to account for the sequential nature of frequencies. In practice, the losses are inversely weighted to the magnitude of the cumulative residual loss from the previous frequencies~\cite{wang2022respecting}. Details are omitted due to lack of space.

The main limitation of the loss in \eqref{eq:basicLoss} is that it can be evaluated only on training data. Following the training strategies used in PINNs, we can use a physical model to regularize the objective evaluated on source locations at any arbitrary resolution in an unsupervised manner \cite{di2022post,wang2021understanding}. One of the main requirements is that our estimated steering vector $\estATF$ and its components are causal for every source position. Anti-causal components may lead to artifacts and incoherent steering vectors. In order to enforce this, we propose a regularization term based on a classic relation saying that the imaginary part of the transfer function must be the Hilbert transform $\mathcal{H}$ of the real part of the transfer function~\cite{papoulis1977signal}:

\begin{equation}
    \mathcal{L}_{\text{Causal}} = \left\| \mathcal{H} \left( \Re \left\{ [ \hat{h}_{ij}]_F \right\} \right)
                                         - \Im \left\{ [\hat{h}_{ij}]_F \right\} \right\|_2^2 ,
\end{equation}
where $\Re \{\cdot \}$ and $\Im \{ \cdot \}$ are the real and imaginary part.
This regularization term can be computed straightforwardly in the DF case by sampling randomly source positions in $\mathbb{R}^3$. 
In the CF condition, input frequencies need to be sampled. To simplify the computation of the Hilbert and Fourier transform, we chose to sample a random number of equally distributed frequencies in $[0,F_s]$.

\section{Evaluation}

\begin{figure}[!t]
    \centering
    \includegraphics[trim={40 0 30 0},clip,width=\linewidth]{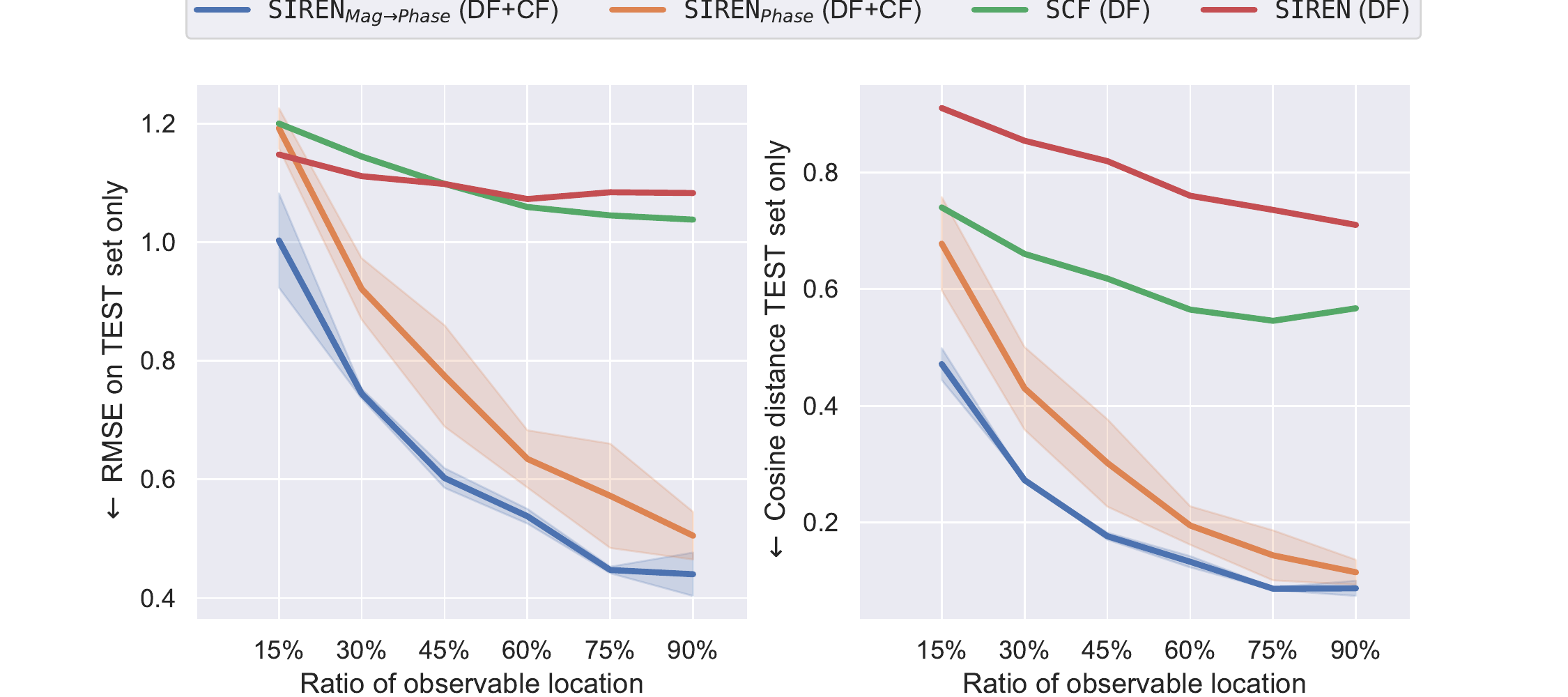}
    \includegraphics[trim={40 0 50 0},clip,width=\linewidth]{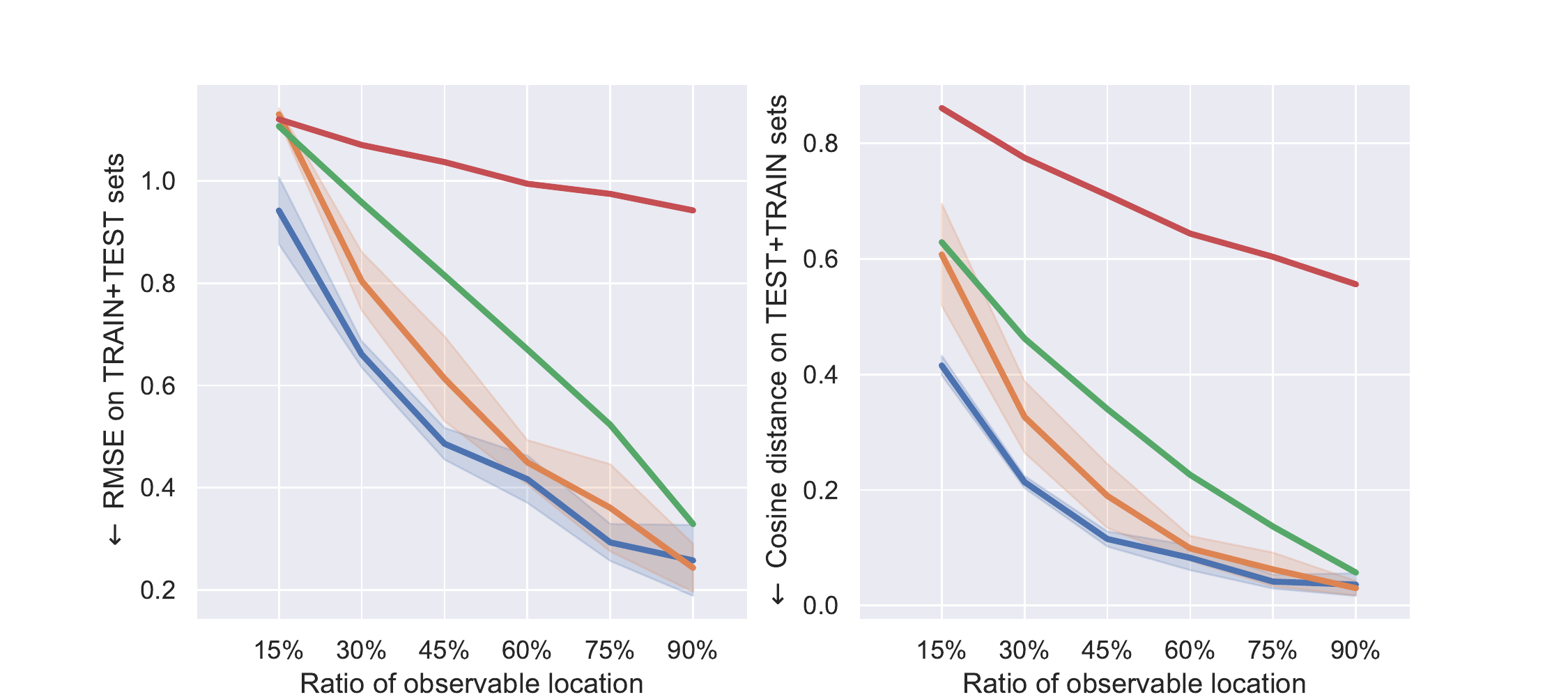}
    \caption{RMSE (left) and cosine distance (right) of time-domain steering vectors for locations in the test set (top) and the whole sphere reconstruction on random data (bottom). Shaded regions show the confidence interval for both continuous and discrete frequency models.}
    \label{fig:exp_results}
    \vspace{-1.8em}
\end{figure}

We aimed to obtain a reliable representation of the steering vectors in both the time and frequency domains.
Therefore, we evaluate the performance of the proposed $\mathtt{SIREN}_{\text{Phase}}$ and $\mathtt{SIREN}_{\text{Mag}\to\text{Phase}}$ models in terms of three metrics, i.e.,
the \textit{root mean square error} (RMSE) and the \textit{cosine distance} 
between the estimated and reference steering vectors in the time domain and
the \textit{log-spectral distance} (LSD) [dB] between the magnitudes of the estimated and reference HRTFs.
We compare our models against a vanilla SIREN as used in~\cite{zhang2022hrtf} ($\mathtt{SIREN}^*$) modified to return magnitude and phase, and an interpolation method based on the spatial characteristic function ($\mathtt{SCF}^*$) \cite{freeland2004interpolation}.

Our evaluation used the EasyCom Dataset~\cite{donley2021easycom} featuring steering vectors of a 6-channel microphone array consisting of 4 microphones attached to head-worn smart glasses and 2 binaural microphones located in the user's ear canals. All data were measured at $F_s = 48 \, \text{kHz}$ on a spherical grid with $60$ equally spaced azimuths and $17$ quasi-equally spaced elevations. If not otherwise specified, we consider $F = 257$ positive frequencies.

Our $\mathtt{SIREN}_{\text{Phase}}$ and $\mathtt{SIREN}_{\text{Mag}\to\text{Phase}}$ models utilized a SIREN architecture composed of four 512-dimensional hidden layers. Additionally, $\mathtt{SIREN}_{\text{Mag}\to\text{Phase}}$ used a second SIREN featuring two 512-dimensional hidden layers. Those models were trained using a batch size of $| \mathcal{B} | = 18$, and a learning rate that is initialized to $10^{-3}$ and scaled by a factor of $0.98$ at every epoch. An early stopping mechanism is applied to avoid overfitting by monitoring a loss computed on a part (20\%) of the training data. In all experiments, we set $\lambda_1 =\lambda_2 = 10$ to match the scale of the different loss terms. $\mathcal{L}_\text{Causal}$ is computed using $B$ random coordinates uniformly sampled in the unit sphere. The baseline $\mathtt{SIREN}^*$ used the same parameterization applied to our $\mathtt{SIREN}_{\text{Phase}}$ model, but output only magnitude and phase directly, without the transformation in \eqref{eq:phase}.

We first considered the task of interpolating the steering vector on a regular grid.
In this task, the training dataset consisted of steering vectors and locations downsampled regularly by a factor of 2 as in~\cite{zhang2022hrtf}. Table~\ref{tab:results} reports the average RMSE and cosine distance of the unseen (missing) data points at full frequency resolution.
Our models trained with the proposed regularizers outperformed the baseline in both metrics. 
Specifically, the performances were improved by introducing new regularization techniques and conditioning the phase estimation on the magnitude.
Additionally, in the context of continuous frequency modeling, incorporating the off-grid regularizer $\mathcal{L}_\text{Causal}$ results in comparable reconstruction results to those achieved through discrete frequency modeling (with a p-value of $0.0032$).

Subsequently, we analyzed the reconstruction capabilities as a function of available points (in percentage) randomly sampled from the original grid.
Fig.~\ref{fig:exp_results} displays the two time-domain objectives for the unseen data only (top) and on the full test dataset (bottom), which comprises training and unseen data points. 
The curves reaffirm the results discussed earlier, showing that conditioning phase estimation on magnitude estimates yields better results.
Notably, it can be observed that the $\mathtt{SIREN}^*$ model trained solely to minimize the $\ell_2$ norm in the complex domain exhibits suboptimal performance in the reconstruction of both seen and unseen points, even when 90$\%$ of the points are seen during training.

\begin{figure}[!t]
    \centering
    \begin{subfigure}[b]{0.48\linewidth}
        \includegraphics[trim={0 0 50 20},clip,width=\linewidth]{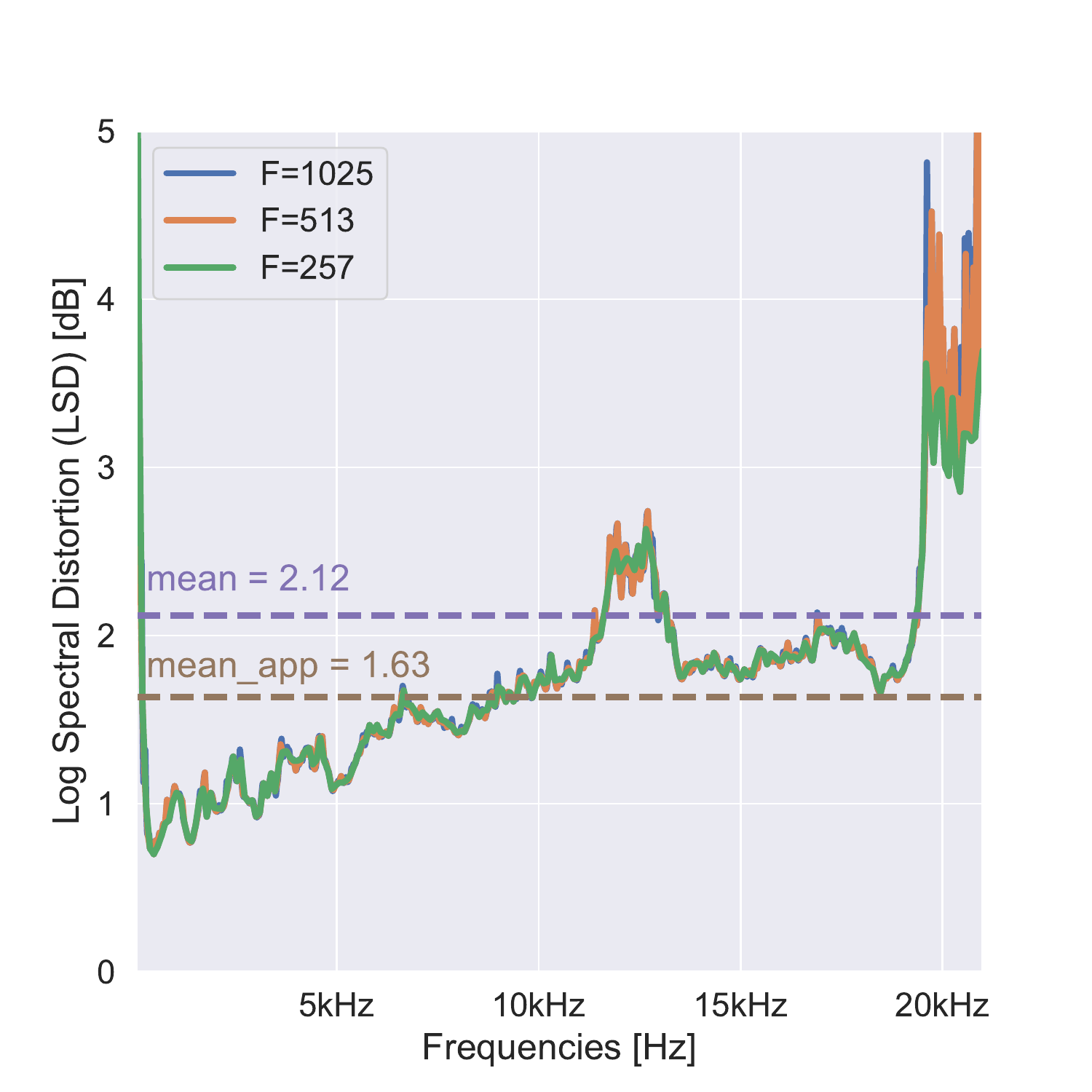}
    \end{subfigure}
     \hfill
     \begin{subfigure}[b]{0.48\linewidth}
        \includegraphics[trim={0 0 50 20},clip,width=\linewidth]{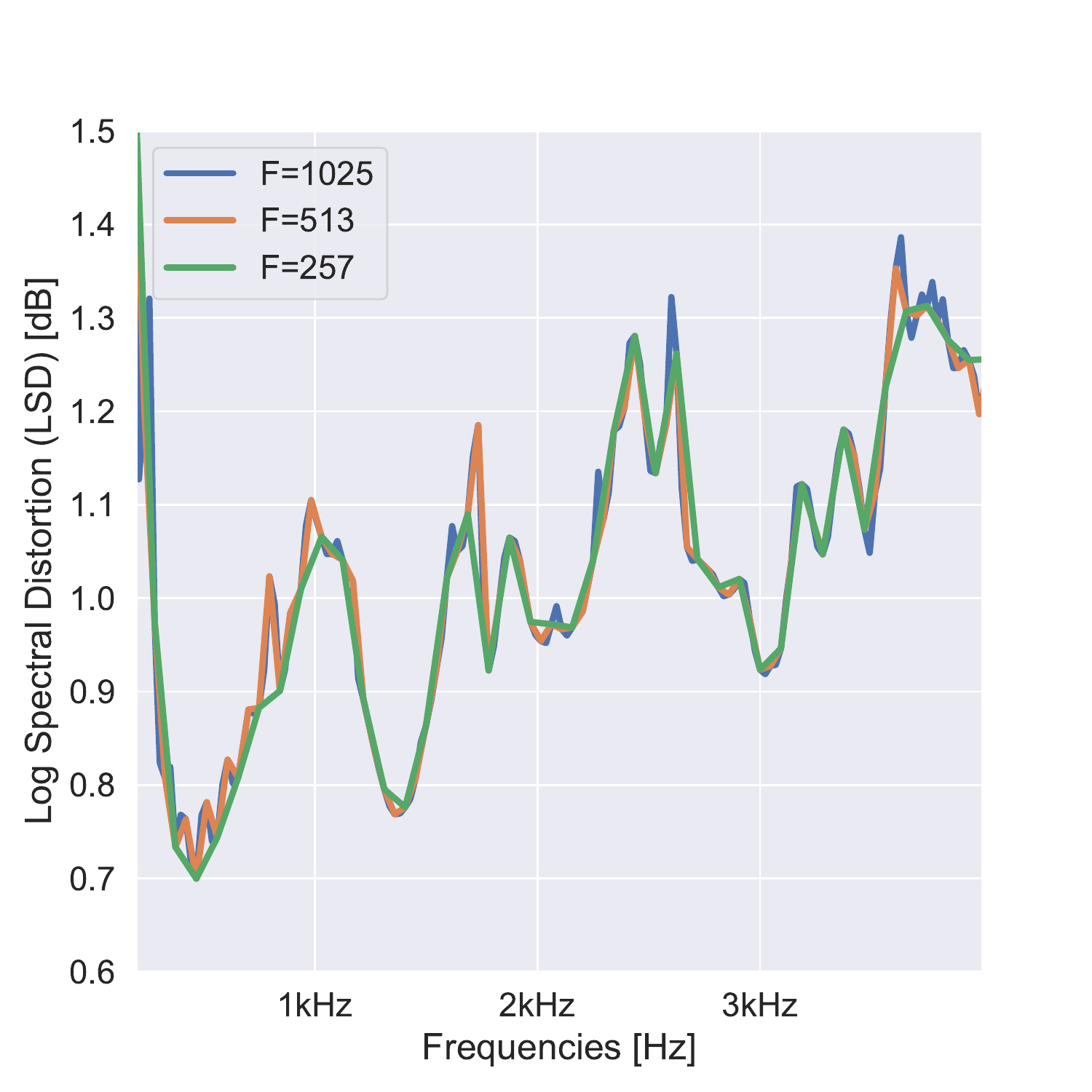}
     \end{subfigure}
    \caption{\protect\label{fig:freq_interp} Average log-spectral distortion at different resolutions in the whole (left) and a selected (right) frequency range. The model was trained on $F=257$ frequency bins.} 
\end{figure}

\begin{table}[!t]
\setlength{\tabcolsep}{4pt}
\caption{\protect\label{tab:results} Average RMSE and cosine distance (in parentheses) between the estimated and reference time-domain steering vectors in the steering vector interpolation task. The values average over the 6 channels. * denotes baseline methods.}
\vspace{-1mm}
\resizebox{\columnwidth}{!}{%
\begin{tabular}{rrc|ccc}
\toprule
Freqs & Model                   & $\mathcal{L}_\text{MSE}$ 
                                & $\mathcal{L}_\text{LogMag} + \lambda_1 \mathcal{L}_\text{Phase}$ 
                                & $+\lambda_2 \mathcal{L}_\text{iDFT}$
                                & $+\mathcal{L}_\text{Causal}$ \\
 \hline
 DF  & $\mathtt{SCF}^*$                                     & 1.42 (0.94) &         -              &            -   &        -         \\
 DF  & $\mathtt{SIREN}^*$                                   & 1.09 (0.78) &         -              & 1.02 (0.70)    &  \textbf{1.01 (0.73)} \\
 \hline
DF   & $\mathtt{SIREN}_\text{Phase}$            & 1.26 (0.77) &   0.36 (0.06)       & 0.34 (0.06)  & \textbf{0.34 (0.06)}  \\
DF   & $\mathtt{SIREN}_{\text{Mag}\to\text{Phase}}$  & 1.22 (0.66) &   0.32 (0.05)       & 0.29 (0.05)  & \textbf{0.28 (0.05)}  \\
 \hline
 CF  & $\mathtt{SIREN}_\text{Phase}$            & 1.42 (0.88) &   0.63 (0.17)       & 0.58 (0.17)   & \textbf{0.33 (0.06)}  \\
 CF  & $\mathtt{SIREN}_{\text{Mag}\to\text{Phase}}$  & 1.37 (0.80) &   0.53 (0.13)       & 0.45 (0.09)   & \textbf{0.37 (0.08)  }\\
\bottomrule
\end{tabular}
}
\vspace{-1.8em}
\end{table} 

Finally, we analyzed the interpolation performances along the frequency axis.
Specifically, we trained a CF variant of $\mathtt{SIREN}_{\text{Mag}\to\text{Phase}}$ model to fit steering vectors measured at $F=257$ frequencies and then evaluated at varying spectral resolutions. 
The results, presented in Fig.~\ref{fig:freq_interp}, demonstrate that the reconstruction error remains consistently low across different evaluation grids. However, it should be noted that the errors increase with the frequency and become more pronounced at the boundaries. This observation could be attributed to the band-limited nature of the target signal. It was also in line with previous findings reported in~\cite{zhang2022hrtf}. Nevertheless, within the frequency range commonly utilized in classical speech processing applications, namely $f \in [40,20000]$ Hz, the mean LSD was around 0.5 dB lower.

\section{Conclusion}
This paper proposes a novel neural field model that provides a \textit{continuous} encoding of steering vectors over both spatial and frequency coordinates given a \textit{discrete} set of measurements. 
Our approach places a strong emphasis on accurately capturing the phase component of the target steering vector while enforcing that the filters maintain their causal nature. 
Our experimental results have illustrated the effectiveness of our model in reconstructing real steering vector measurements.
As we look ahead, we envision a wide array of potential applications. Specifically, the continuous synthesis of steering vectors holds promise for tasks such as beamforming in sound source localization and separation. Furthermore, the versatility of our proposed model extends to microphone calibration, as we optimize microphone positions as model parameters.

\begingroup
\newcommand{\myfontsize}{\fontsize{9.0}{10.0}\selectfont}
\def\baselinestretch{.96}\let\normalsize\myfontsize\normalsize

\bibliographystyle{IEEEbib}         
\bibliography{IEEEabrv,shorterMYabrv,references}
\endgroup

\end{document}

%% file: notations.tex
\newcommand{\srcPos}{\ensuremath{\mathbf{s}}}
\newcommand{\micPos}{\ensuremath{\mathbf{m}}}
\newcommand{\refPos}{\ensuremath{\mathbf{r}}}

\newcommand{\az}{\ensuremath{\theta}}
\newcommand{\el}{\ensuremath{\varphi}}
\newcommand{\model}{\mathcal{M}}

\newcommand{\estATF}{\ensuremath{\hat{h}_{ij}}}
\newcommand{\refATF}{\ensuremath{h_{ij}}}

%% file: main.bbl
\begin{thebibliography}{10}

\bibitem{donley2021easycom}
Jacob Donley, Vladimir Tourbabin, Jung-Suk Lee, Mark Broyles, Hao Jiang, Jie Shen, Maja Pantic, Vamsi~Krishna Ithapu, and Ravish Mehra,
\newblock ``{EasyCom}: An augmented reality dataset to support algorithms for easy communication in noisy environments,'' arXiv e-print, 2021,
\newblock arXiv:2107.04174v2.

\bibitem{sekiguchi2020fastmnmf}
Kouhei Sekiguchi, Yoshiaki Bando, Aditya~Arie Nugraha, Kazuyoshi Yoshii, and Tatsuya Kawahara,
\newblock ``Fast multichannel nonnegative matrix factorization with directivity-aware jointly-diagonalizable spatial covariance matrices for blind source separation,''
\newblock {\em {IEEE/ACM} Trans. Audio, Speech, Language Process.}, vol. 28, pp. 2610--2625, 2020.

\bibitem{schmidt1986multiple}
Ralph Schmidt,
\newblock ``Multiple emitter location and signal parameter estimation,''
\newblock {\em {IEEE} Trans. Antennas Propag.}, vol. 34, no. 3, pp. 276--280, 1986.

\bibitem{annibale2012geometric}
Paolo Annibale, Jason Filos, Patrick~A Naylor, and Rudolf Rabenstein,
\newblock ``Geometric inference of the room geometry under temperature variations,''
\newblock in {\em Proc. Int. Symp. Control Commmun. Signal Process.}, 2012, pp. 1--4.

\bibitem{gannot2001signal}
Sharon Gannot, David Burshtein, and Ehud Weinstein,
\newblock ``Signal enhancement using beamforming and nonstationarity with applications to speech,''
\newblock {\em {IEEE} Trans. Signal Process.}, vol. 49, no. 8, pp. 1614--1626, 2001.

\bibitem{argentieri2015survey}
Sylvain Argentieri, Patrick Danes, and Philippe Sou{\`e}res,
\newblock ``A survey on sound source localization in robotics: From binaural to array processing methods,''
\newblock {\em Computer Speech \& Language}, vol. 34, no. 1, pp. 87--112, 2015.

\bibitem{pulkki1997virtual}
Ville Pulkki,
\newblock ``Virtual sound source positioning using vector base amplitude panning,''
\newblock {\em Journal of the audio engineering society}, vol. 45, no. 6, pp. 456--466, 1997.

\bibitem{freeland2004interpolation}
Fábio~P. Freeland, Luiz W.~P. Biscainho, and Paulo S.~R. Diniz,
\newblock ``Interpolation of head-related transfer functions ({HRTFS}): A multi-source approach,''
\newblock in {\em Proc. EUSIPCO}, 2004.

\bibitem{zotkin2009regularized}
Dmitry~N. Zotkin, Ramani Duraiswami, and Nail~A. Gumerov,
\newblock ``Regularized {HRTF} fitting using spherical harmonics,''
\newblock in {\em Proc. {IEEE} WASPAA}, 2009, pp. 257--260.

\bibitem{tancik2020fourier}
Matthew Tancik, Pratul Srinivasan, Ben Mildenhall, Sara Fridovich-Keil, Nithin Raghavan, Utkarsh Singhal, Ravi Ramamoorthi, Jonathan Barron, and Ren Ng,
\newblock ``Fourier features let networks learn high frequency functions in low dimensional domains,''
\newblock {\em Proc. NeurIPS}, vol. 33, pp. 7537--7547, 2020.

\bibitem{sitzmann2020implicit}
Vincent Sitzmann, Julien Martel, Alexander Bergman, David Lindell, and Gordon Wetzstein,
\newblock ``Implicit neural representations with periodic activation functions,''
\newblock in {\em Proc. NeurIPS}, 2020, vol.~33, pp. 7462--7473.

\bibitem{xie2022neural}
Yiheng Xie, Towaki Takikawa, Shunsuke Saito, Or~Litany, Shiqin Yan, Numair Khan, Federico Tombari, James Tompkin, Vincent Sitzmann, and Srinath Sridhar,
\newblock ``Neural fields in visual computing and beyond,''
\newblock in {\em Comput. Graph. Forum}, 2022, vol.~41, pp. 641--676.

\bibitem{mildenhall2020nerf}
Ben Mildenhall, Pratul~P. Srinivasan, Matthew Tancik, Jonathan~T. Barron, Ravi Ramamoorthi, and Ren Ng,
\newblock ``{NeRF}: Representing scenes as neural radiance fields for view synthesis,''
\newblock in {\em Proc. ECCV}, 2020, pp. 405--421.

\bibitem{gafni2021dynamic}
Guy Gafni, Justus Thies, Michael Zollhofer, and Matthias Nie{\ss}ner,
\newblock ``Dynamic neural radiance fields for monocular {4D} facial avatar reconstruction,''
\newblock in {\em Proc. IEEE/CVF Conf. Comput. Vis. Pattern Recog.}, 2021, pp. 8649--8658.

\bibitem{raissi2019physics}
Maziar Raissi, Paris Perdikaris, and George~E. Karniadakis,
\newblock ``Physics-informed neural networks: A deep learning framework for solving forward and inverse problems involving nonlinear partial differential equations,''
\newblock {\em J. Comput. Phys.}, 2019.

\bibitem{karniadakis2021physics}
George~Em Karniadakis, Ioannis~G Kevrekidis, Lu~Lu, Paris Perdikaris, Sifan Wang, and Liu Yang,
\newblock ``Physics-informed machine learning,''
\newblock {\em Nature Reviews Phys.}, vol. 3, no. 6, pp. 422--440, 2021.

\bibitem{wang2021understanding}
Sifan Wang, Yujun Teng, and Paris Perdikaris,
\newblock ``Understanding and mitigating gradient flow pathologies in physics-informed neural networks,''
\newblock {\em SIAM J. Sci. Comput.}, vol. 43, no. 5, pp. A3055--A3081, 2021.

\bibitem{richard2022deep}
Alexander Richard, Peter Dodds, and Vamsi~Krishna Ithapu,
\newblock ``Deep impulse responses: Estimating and parameterizing filters with deep networks,''
\newblock in {\em Proc. {IEEE} ICASSP}, 2022.

\bibitem{zhang2022hrtf}
You Zhang, Yuxiang Wang, and Zhiyao Duan,
\newblock ``{HRTF} field: Unifying measured {HRTF} magnitude representation with neural fields,''
\newblock in {\em Proc. {IEEE} ICASSP}, 2023.

\bibitem{lee2022global}
Jin~Woo Lee, Sungho Lee, and Kyogu Lee,
\newblock ``Global {HRTF} interpolation via learned affine transformation of hyper-conditioned features,''
\newblock in {\em Proc. {IEEE} ICASSP}, 2023.

\bibitem{lee2022neural}
Jin~Woo Lee and Kyogu Lee,
\newblock ``Neural fourier shift for binaural speech rendering,''
\newblock in {\em Proc. {IEEE} ICASSP}, 2023.

\bibitem{luo2022learning}
Andrew Luo, Yilun Du, Michael~J Tarr, Joshua~B Tenenbaum, Antonio Torralba, and Chuang Gan,
\newblock ``Learning neural acoustic fields,''
\newblock in {\em Proc. NeurIPS}, 2022, pp. 1--13.

\bibitem{su2022inras}
Kun Su, Mingfei Chen, and Eli Shlizerman,
\newblock ``Inras: Implicit neural representation for audio scenes,''
\newblock {\em NeurIPS}, 2022.

\bibitem{pezzoli2023implicit}
Mirco Pezzoli, Fabio Antonacci, and Augusto Sarti,
\newblock ``Implicit neural representation with physics-informed neural networks for the reconstruction of the early part of room impulse responses,''
\newblock in {\em Proc. Forum Acousticum.}, 2023.

\bibitem{ma2023physics}
Fei Ma, Thushara~D Abhayapala, Prasanga~N Samarasinghe, and Xingyu Chen,
\newblock ``Physics informed neural network for head-related transfer function upsampling,''
\newblock {\em arXiv preprint arXiv:2307.14650}, 2023.

\bibitem{vincent2018audio}
E.~Vincent, T.~Virtanen, and S.~Gannot,
\newblock {\em Audio Source Separation and Speech Enhancement},
\newblock John Wiley \& Sons, 2018.

\bibitem{scheibler2018pyroomacoustics}
Robin Scheibler, Eric Bezzam, and Ivan Dokmani{\'c},
\newblock ``Pyroomacoustics: A python package for audio room simulation and array processing algorithms,''
\newblock in {\em Proc. {IEEE} ICASSP}, 2018.

\bibitem{srivastava2023virtually}
Prerak Srivastava, Antoine Deleforge, Archontis Politis, and Emmanuel Vincent,
\newblock ``How to (virtually) train your speaker localizer,''
\newblock in {\em INTERSPEECH 2023}, 2023.

\bibitem{nugraha19dgm}
Aditya~Arie Nugraha, Kouhei Sekiguchi, and Kazuyoshi Yoshii,
\newblock ``A deep generative model of speech complex spectrograms,''
\newblock in {\em Proc. {IEEE} ICASSP}, 2019, pp. 905--909.

\bibitem{richard2021neural}
Alexander Richard, Dejan Markovic, Israel~D. Gebru, Steven Krenn, Gladstone~Alexander Butler, Fernando Torre, and Yaser Sheikh,
\newblock ``Neural synthesis of binaural speech from mono audio,''
\newblock in {\em Proc. ICLR}, 2021, pp. 1--13.

\bibitem{donley2022dare}
Jacob Donley and Paul Calamia,
\newblock ``{DARE-Net}: Speech dereverberation and room impulse response estimation,''
\newblock Tech. {R}ep., Stanford University, 2022.

\bibitem{wang2022respecting}
Sifan Wang, Shyam Sankaran, and Paris Perdikaris,
\newblock ``Respecting causality is all you need for training physics-informed neural networks,'' arXiv e-print, 2022,
\newblock arXiv:2203.07404v1.

\bibitem{di2022post}
Diego Di~Carlo, Dominique Heitz, and Thomas Corpetti,
\newblock ``Post processing sparse and instantaneous {2D} velocity fields using physics-informed neural networks,''
\newblock in {\em Proc. Int. Symp. Appl. Laser Imag. Tech. Fluid Mech.}, 2022.

\bibitem{papoulis1977signal}
Athanasios Papoulis,
\newblock {\em Signal analysis},
\newblock Mcgraw-Hill College, 1977.

\end{thebibliography}
